\documentclass[
 reprint,
 amsmath,amssymb,
 aps,
pra,
floatfix,
]{revtex4-1}
\usepackage[utf8x]{inputenc}
\usepackage{natbib}
\usepackage[T2A]{fontenc}
\usepackage{graphicx,lipsum}
\usepackage{dcolumn}
\usepackage{bm}
\usepackage[dvipsnames]{xcolor}
\usepackage{array,multirow}
\usepackage{bbm}
\usepackage{amsmath}
\usepackage{epstopdf}
\UseRawInputEncoding

\begin{document}

\preprint{APS/123-QED}
\title{ Quantum speed of evolution of  neutral mesons}

\author{Subhashish Banerjee\textsuperscript{}}
\altaffiliation[subhashish@iitj.ac.in]{}
\author{K. G. Paulson\textsuperscript{}}
\email{paulsonkgeorg@gmail.com}
\affiliation{Indian Institute of Technology Jodhpur, Jodhpur-342011, India}



\date{\today}

\begin{abstract}
We investigate the quantum-mechanical time-evolution speed limit for neutral   $K$ and $B$ mesons, both single as well as correlated, within the framework of open quantum systems. The role of coherence--mixing, a crucial feature of the open system evolution of the underlying quantum systems (here, the mesons), on the quantum-mechanical time-evolution speed limit is studied. The impact of decoherence and CP (charge conjugation parity) symmetry violation on quantum-mechanical time-evolution speed limit is also investigated.  The quantum-mechanical time-evolution speed limit increases with the evolution time for the single mesons, a signature of the underlying open system dynamics of the evolution being semi-group in nature. The evolution of the correlated mesons slows down for an evolution time of approximately one-fourth of the lifetime, after which it is sped up. An overall pattern that emerges is that correlated mesons evolve faster, as compared to their uncorrelated counterparts, suggesting that quantum correlations can speed up evolution.
\end{abstract}

\maketitle


\section{\label{sec:level1} Introduction } 
The theory of quantum mechanics has revolutionized our understanding of the universe at the smallest scales. One of the key features of quantum mechanics is that it allows for the existence of superposition and entanglement. Significant advancements in science and technology have been made possible by exploiting entanglement and superposition, which would be otherwise impossible. This has inspired the investigation of quantum correlations and their applications in high energy systems also, in particular, meson and neutrino systems~\cite{2,3,4,5,6,8,11,12}. The~nature of spatial~\cite{brabon,blassone,3flavor,2flavor,sbbani,kerbikov,dajka,dixit2019} and temporal~\cite{EntropicLG,LG3Flavor,LGMesons,LG3FlavorDetail,LGsubatomic} quantum correlations, quantum coherence~\cite{sbkhushboo}, geometric phase~\cite{sbgpneutrino,bani,poonam,luczka,vitiello,fuller,lu2021}, non-standard interactions~\cite{aloktripta}, as~well as the impact of decoherence in neutral meson and neutrinos, have also been studied~\cite{mavrapolis,sbalok2014,banerjee2016quantum,sin2beta,ellis,peskin,sarkar}.

Heisenberg's uncertainty principle is a fundamental concept in quantum mechanics. The uncertainty principle for position and momentum shows the impossibility of simultaneous measurement of both position and momentum precisely, but the interpretation of  the  energy-time uncertainty relation in this line is not so obvious.  In~\cite{mandestamm},  Mandelstam and Tamm (MT) showed that the energy-time uncertainty principle ($\Delta t\gtrsim\frac{\hbar}{\Delta E}$) is a statement about the intrinsic time in which a quantum system evolves. The notion of the speed of transportation of a quantum system, the~quantum-mechanical time-evolution speed limit (QMTSL), was first introduced in~\cite{anandan} using the geometric approach, making use of the Fubini-Study metric on the space of quantum pure states. The significance of this is that the length of any path connecting two distinguishable pure states is lower bounded by the geodesic length between them according to the Fubini-Study metric.
Margolus, Levitin  (ML)~\cite{Margolous} provided a different QMTSL, which depends on the mean energy. 
The QMTSL for unitary dynamics, restricted to orthogonal pure states, can be made tighter by combining the MT and ML bounds as $\tau_{QSL}=\max\Bigg\{\frac{\pi\hbar}{2\Delta H},\frac{\pi\hbar}{2\langle H\rangle-E_{0}}\Bigg\}$. $\tau_{QSL}$ refers to the minimum amount of time required for a quantum system to evolve from one state to another.  It sets a fundamental limit on the speed at which quantum information can be processed or transmitted and is also connected to the stability of matter~\cite{sbiiscqsl}. It plays a crucial role in the development of quantum technologies such as quantum computing and quantum communication. Any real system interacts with its environment, which can lead to a loss of quantum coherence. This can be taken into account by using the ideas of open quantum systems~\cite{breuerpettrucione,caban2005unstable,sbbook}.  The~extension of the QMTSL to open quantum systems~\cite{deffner2013, adcampo2013, deffner2017, pires2016} has important implications for the design and implementation of quantum information processing protocols~\cite{deffner2017}. Recently, a good amount of work has been done in this domain, its connection with resource theory, and~usefulness for other technological  aspects~\cite{pires2016, actionqsl, baruahqsl, wei2016,paulson2021hierarchy,paulson2022,sbcentralspinspeed}.

The treatment of unstable systems has a long history~\cite{21,24,25,26,28,29,30,32,33,34}.
The study is facilitated by the use of the density matrix formalism, a~view supported by work on decaying systems using dynamical semi-groups~\cite{35}, where it was suggested that a unitary evolution of an unstable system along with its decay products is not feasible. Hence, a~decaying system is intrinsically an open system, even without explicitly invoking an external environment.

In this work, we aim to investigate the role of the QMTSL in the dynamics of the decaying single and the correlated neutral mesons, $K^0$, $B^0_d$ and $B^0_s$. The neutral mesons exhibit the interesting feature of flavour oscillations. A study of quantum speed would thus serve to develop insight into the dynamics of flavour oscillations. Further, since mesons are decaying systems, the study also serves to bring out the features of quantum speed in the context of unstable quantum systems. Quantum coherence plays a crucial role in our understanding of evolution. Decaying meson evolution, being inherently open, leads to mixing. It would thus be meaningful to study the impact of coherence and mixing~\cite{singh2015maximally} on the quantum speed of the meson dynamics.
Quantum correlations are the epitome of quantumness in a system and highlight the utility of its quantum nature~\cite{horodecki,patarek,gisin,briegel1,briegel2}. An~interplay between various facets of quantum correlations and the CP-symmetry (charge conjugation parity symmetry) violation in the quark sector has been actively studied~\cite{banerjee2016quantum,bramon,domenico}.
 Violation in the quark sector has been shown to be a crucial ingredient in violating a Bell inequality for entangled $K$-meson pairs~\cite{8}.  Another fact bolstering the importance of CP symmetry violation in particle physics is the parameter $\sin 2\beta$, whose measurement was the first signal for CP violation outside the neutral kaon system and helped to establish the CKM mechanism for CP  violation. Correlated neutral $K^0$ and $B^0$ mesons are obtained from the decay and subsequent hadronization of the $\phi$ and $\Upsilon$ resonances, respectively. Analysis of correlated neutral mesons from the perspective of quantum speed would thus be pertinent. The role of decoherence and $CP$ (charge-parity) violation are also considered.

The plan of the paper is as follows. We provide  information related to the QMTSL and the notion of coherence and mixing in Section~\ref{pre}. This is followed by a brief discussion of both single and correlated neutral mesons and  the analytical expressions of coherence and entropy for these systems (Section~\ref{NM}). In~Section~\ref{results}, these systems' QMTSL is presented and discussed, followed by the concluding remarks in Section~\ref{conclusion}.

\section{Preliminaries}\label{pre}
The basic notion of the QMTSL and coherence-mixing balance are presented below.
\subsection{Measures of Quantum-Mechanical Time-Evolution Speed~Limit}
Mandelstam-Tamm and Margolus-Levitin-types bound on speed limit time for an  
open quantum system based on the geometrical distance between the initial pure state $\rho_{0}=\vert\psi_{0}\rangle\langle\psi_{0}\vert$ and final state $\rho_{\tau}$, which is a tighter bound is given as~\cite{deffner2013},
\begin{equation}
    \tau_{QSL}=\max\Bigg\{\frac{1}{\Lambda^{\textrm{op}}_{\tau}},\frac{1}{\Lambda^{\textrm{tr}}_{\tau}},\frac{1}{\Lambda^{\textrm{hs}}_{\tau}}\Bigg\} \sin^2[\mathcal{B}],
    \label{spdlmt}
\end{equation}
where $\frac{1}{\Lambda^{\textrm{op}}_{\tau}}$,$\frac{1}{\Lambda^{\textrm{tr}}_{\tau}}$, and~$\frac{1}{\Lambda^{\textrm{hs}}_{\tau}}$ corresponds to the operator, trace, and~Hilbert-Schmidt norms, respectively. Bures angle $\mathcal{B}(\rho_{0},\rho_{\tau})=\arccos\sqrt{\mathcal{F}(\rho_{0},\rho_{\tau})}$, where the Bures fidelity $\mathcal{F}(\rho_{0},\rho_{\tau})$ is $\bigg[\textrm{tr}[\sqrt{\sqrt{\rho_{0}}\rho_{\tau}\sqrt{\rho_{0}}}]\bigg]^{2}$ and,
\begin{equation}
    \Lambda^\textrm{op,tr,hs}_{\tau}=\frac{1}{\tau}\int^{\tau}_{0}dt\vert\vert \mathcal{L}(\rho_t)\vert\vert_\textrm{op,tr,hs}.
    \label{B_speed_limit}
\end{equation}
As  it is known, the~operators obey the following inequality
$\vert\vert \cdot\vert\vert_{\textrm{op}}\leq\vert\vert \cdot\vert\vert_{\textrm{hs}}\leq\vert\vert \cdot\vert\vert_{\textrm{tr}}$, as~a result, we have, $1/\Lambda^{\textrm{op}}_{\tau}\geq1/\Lambda^{\textrm{hs}}_{\tau}\geq1/\Lambda^{\textrm{tr}}_{\tau}$, which shows that the
QMTSL, based on the operator norm of the nonunitary generator, provides the tighter bound on $\tau_{QSL}$ for an actual evolution time $\tau$. 

The geometrical approach for the quantum speed limit, as discussed here, relies on the geodesic distance between any two points is the shortest possible lengths connecting them. This idea was brought out in~\cite{anandan} in the context of an extension of the MT bound to time-dependent Hamiltonians on the space of quantum pure states. The~geometric quantum speed limits are not sensitive to instantaneous speed; the problem is resolved by the action $\tau_{QSL}$~\cite{actionqsl}. In~\cite{pires2016}, an~important observation was highlighted in that there exist an infinite family of contractive Riemannian metrics on the set of density matrices, which could serve as a geometric quantifier of distinguishability between the states, giving rise to a corresponding quantum speed limit (QSL). In particular, the role of the Wigner--Yanase skew information metric for mixed states was brought out.

\subsection{Coherence and~Mixing }
Quantum coherence, a~quantum resource, is a significant characteristic arising from the superposition of states. Quantification of quantum coherence is dependent on the reference basis considered. Different measures are adopted to quantify the coherence of quantum states. Here, we use the $l_1$ norm measure of quantum coherence, defined for a state $\rho$ and the reference basis $\{\vert i\rangle\}$
\begin{equation}
    C_{l1}(\rho)=\sum_{i\neq j}\vert\rho_{i,j}\vert.
    \label{coh}
\end{equation}
We  have $\rho_{ij}=\langle i\vert\rho\vert j\rangle$.

A quantum system's inescapable interaction with its surroundings  results in the loss of purity. Linear entropy quantifies the mixedness of quantum states due to, for example, their environment. For~a general $d$-dimensional state, linear entropy is defined as
\begin{equation}
    S_{L}(\rho)=\frac{d}{d-1}(1-\textrm{Tr}\rho^2).\label{linearentropy}
    \end{equation}
For a mixed state $0<S_{L}(\rho)\leq 1$. Trade-off between quantum coherence and mixedness ($ M_{Cl} $) for a quantum state $\rho$ is given by the inequality~\cite{singh2015maximally,sbsamya}
\begin{equation}
    S_{L}(\rho)+\frac{C_{l1}^2}{(d-1)^2}\leq1.\label{cohmixing}
\end{equation}
For a maximally coherent mixed state, the above quantity will equal~1.

\section{Neutral~Mesons}\label{NM}

Open quantum system formalism can be made use of to analyse the decay of unstable quantum systems. We consider single  and correlated $B^0~(K^0)$ neutral meson systems.
Decoherence, which is due to the decaying nature of the system, is modelled here by a single phenomenological parameter~\cite{sin2beta} which represents the interaction between the one-particle system and its environment. The~the environment could be attributed to quantum gravity effects~\cite{nano,mavro} or due to the detector background itself.
Apart from decoherence, we also consider the effects of $CP$ violation.

The experiments on the $B_0(K_0)$ meson systems involve the determination of their flavour at the time of production or decay, which is achieved by analyzing the flavour-specific decays.
The charge of the final state lepton in the semileptonic decays of a neutral meson usually determines the flavour of that meson at the time of decay. This is a consequence of the $\Delta B = \Delta Q$ rule for the semileptonic decays of $B$ mesons and is usually assumed in experiments. The process of determination of the initial flavour of a neutral meson is called tagging. This is achieved by making
use of the rule of the associated production. The~mesons are produced either by strong or electromagnetic interactions; hence, a~quark is always produced in association with its anti-quark, as~the flavour is conserved in these interactions. If two entangled neutral mesons are produced, then detecting the flavour-specific final state of one meson at a particular time determines the flavour of that meson as well
as the other meson at that time. The~oscillation probability of the tagged meson is then determined by identifying
its final flavour-specific~state.

\subsection{Single neutral mesons}


Here, we discuss the time decay of $B^0~ (K^0)$ mesons. The Hilbert space for the decaying meson system is ${\mathcal H}_{B^0 (K^0)}\oplus {\mathcal H}_0$,  where ${\mathcal H}_0$ is for vacuum state $\vert0\rangle$ to  incorporate the decay in this system. The orthonormal spanning vectors for $B^0$ meson system are

\begin{widetext}

\begin{figure}[!h]
\centering
\includegraphics[height=45mm,width=0.32\textwidth]{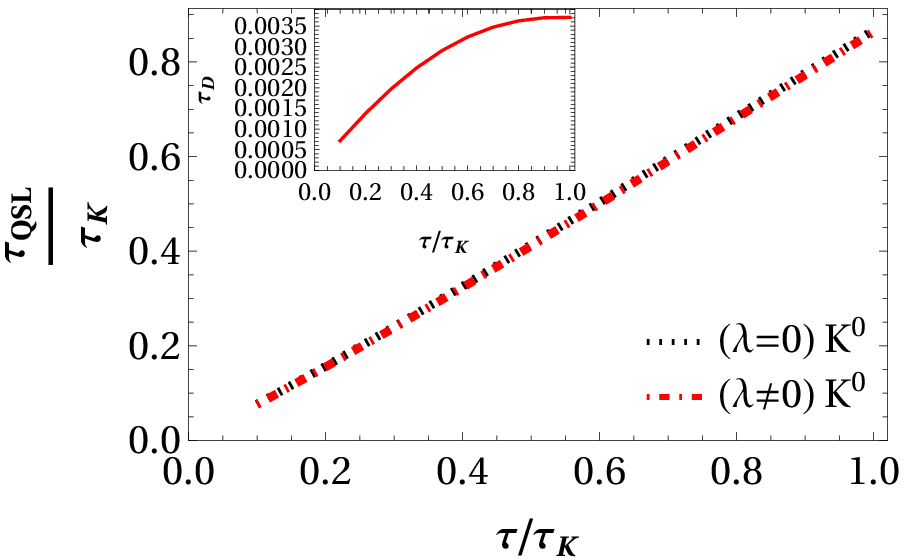}
\includegraphics[height=45mm,width=0.32\textwidth]{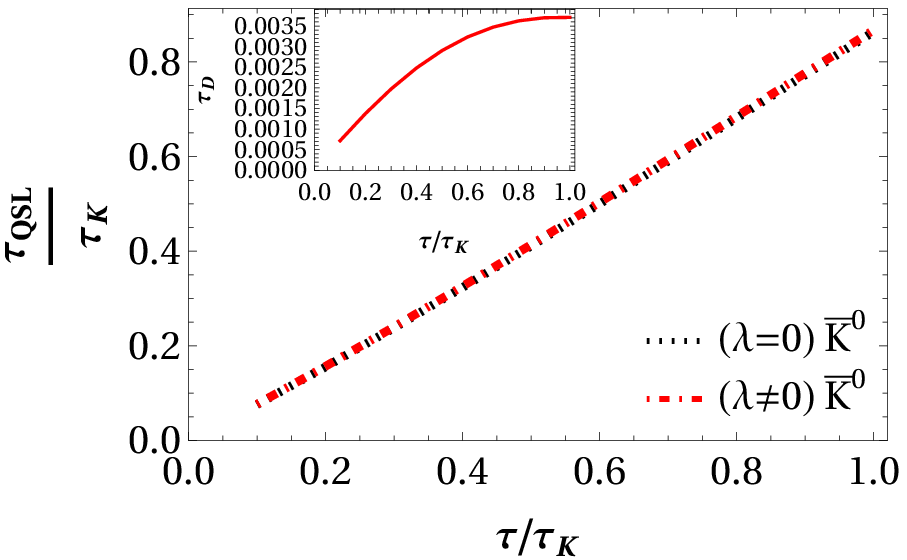}
\includegraphics[height=45mm,width=0.32\textwidth]{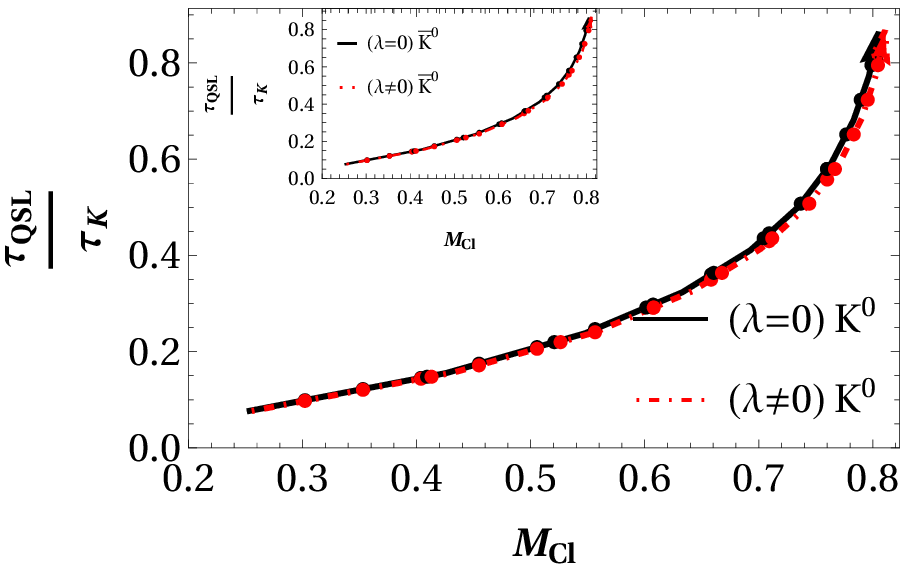}
\caption{QMTSL in terms of Bures measure for the decay of the $K^0$ and $\bar{K}^0$ meson system with and without decoherence effects (left and central subplots, respectively). For  the $K^0$ meson system, the mean
lifetime is $\tau_{K}=1.7889 \times 10^{-10} s$, $\Gamma=5.59\times 10^9 s^{-1}$, $\Delta\Gamma=1.1174\times 10^{10} s^{-1}$, $\Delta m=5.302\times 10^9 s^{-1}$, $m=7.532\times 10^{23} s^{-1}$, $\lambda=2.0\times10^8 s^{-1}$. The insets of the plots highlight the difference caused by taking decoherence, parameterized by $\lambda$, into account. The impact of the coherence-mixedness trade-off on QMTSL is also shown (right subplot).}
    \label{k_meson}
\end{figure}

\begin{figure}[!htb]
    \centering
    \includegraphics[height=45mm,width=0.32\textwidth]{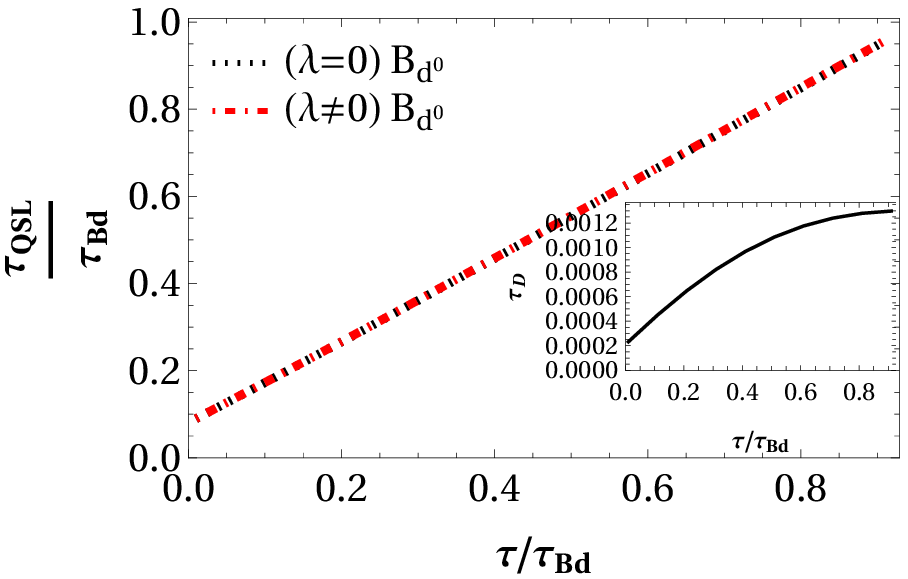}
\includegraphics[height=45mm,width=0.32\textwidth]{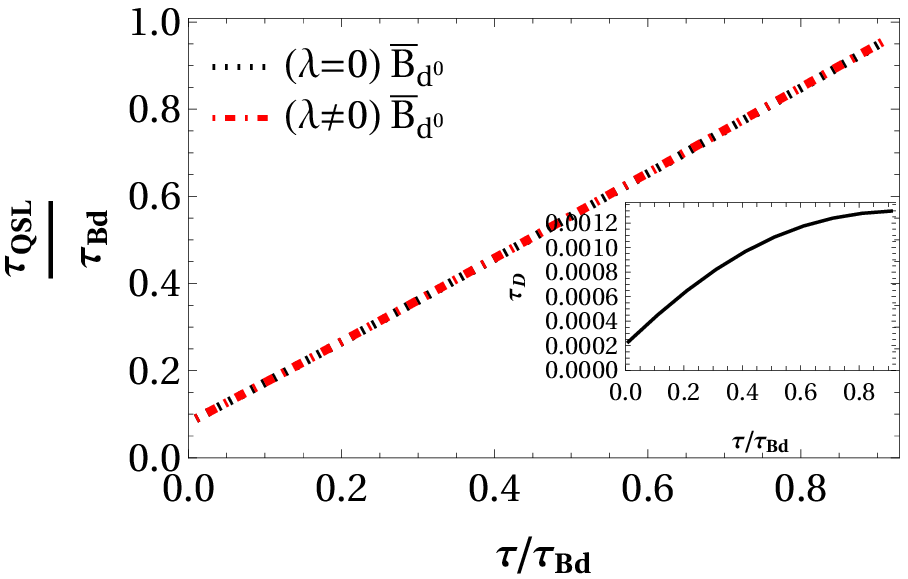}
    \includegraphics[height=45mm,width=0.32\textwidth]{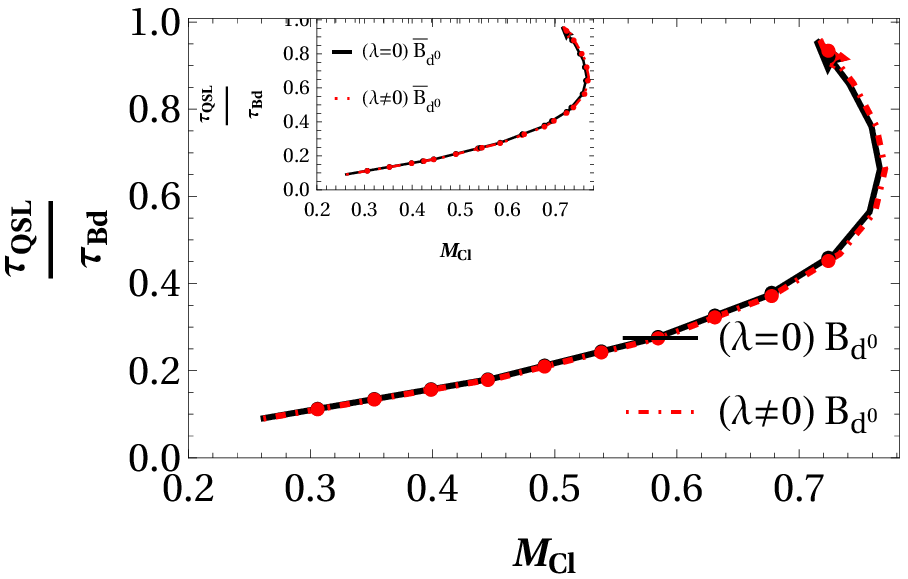}
    \caption{QMTSL in terms of Bures measure for the decay of the $B^0_d$ and $\bar{B}^0_d$ meson system with and without decoherence effects (left and central subplots, respectively). For  the $B^0_d$ meson system, the mean lifetime is  $\tau_{B_d}=1.518 \times 10^{-12} s$, $\Gamma=6.58\times 10^11 s^{-1}$, $\Delta\Gamma=0 s^{-1}$, $\Delta m=0.5064\times 10^12 s^{-1}$,  $m=7.9917 \times 10^{24} s^{-1}$, $\lambda=0.012 \times 10^{12} s^{-1}$. The inset of the plots highlights the difference caused by taking decoherence, parameterized by $\lambda$, into account. The impact of the coherence-mixedness trade-off on QMTSL is also shown (right subplot).}
    \label{bd_meson}
\end{figure}

\begin{figure} [!htb]
    \centering
\includegraphics[height=45mm,width=0.32\textwidth]{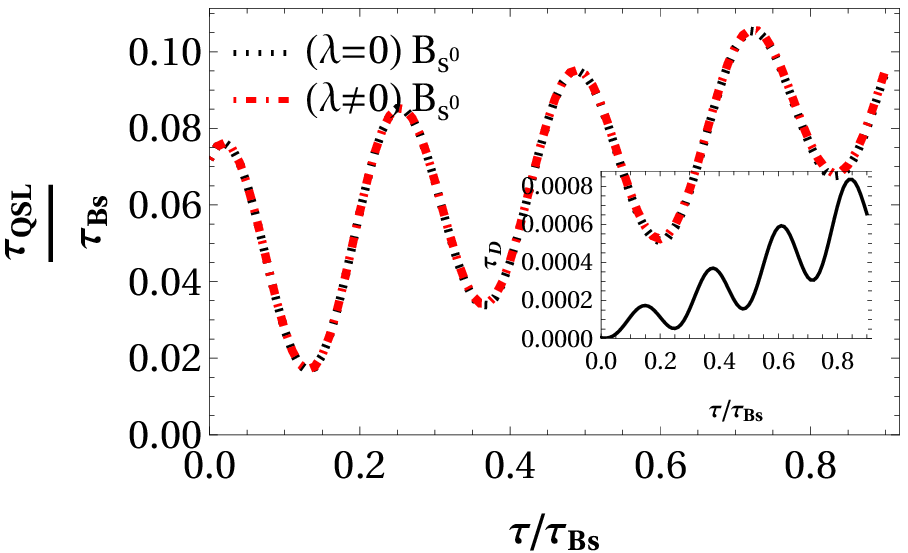}
\includegraphics[height=45mm,width=0.32\textwidth]{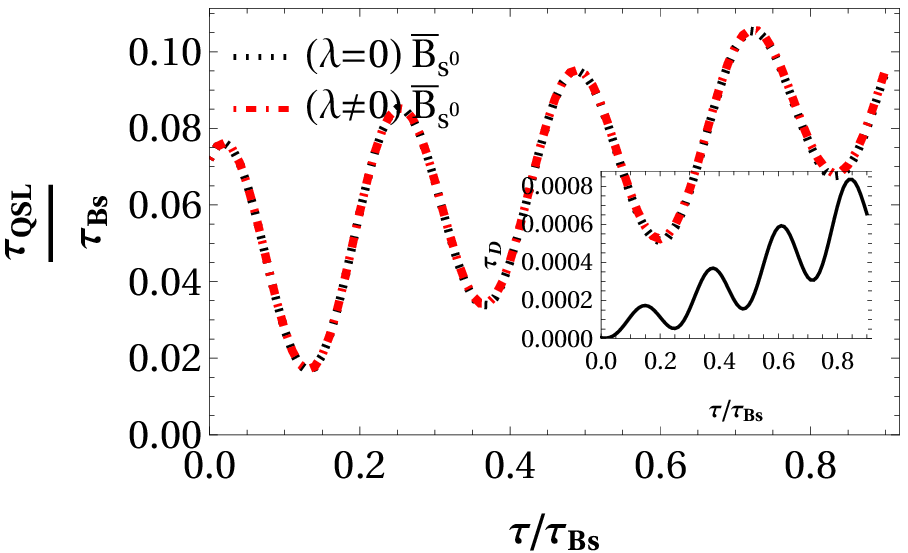}
\includegraphics[height=45mm,width=0.32\textwidth]{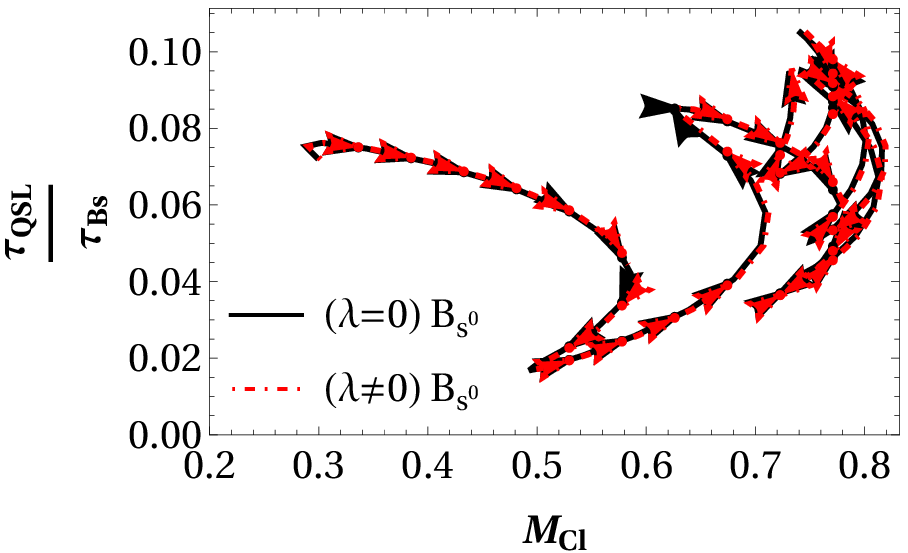}
\caption{QMTSL in terms of Bures measure for the decay of the $B^0_s$ and $\bar{B}^0_s$ meson system with and without decoherence effects (left and central subplots, respectively). For  the $B^0_s$ meson system, the mean lifetime is $\tau_{B_s}=1.509 \times 10^{-12} s$, $\Gamma=0.6645\times 10^{12} s^{-1}$, $\Delta\Gamma=0.086\times 10^{12} s^{-1}$, $\Delta m=17.757\times 10^{12} s^{-1}$, $m=8.123\times 10^{24} s^{-1}$, $\lambda=0.012\times10^{12} s^{-1}$. The inset of the plots highlights the difference caused by taking decoherence, parameterized by $\lambda$, into account. The impact of the coherence-mixedness trade-off on QMTSL is also shown (right subplot).}
    \label{bs_meson}
\end{figure}
\end{widetext}

\begin{equation}
    \vert B^{0}\rangle=\begin{pmatrix}
    1\\0\\0
    \end{pmatrix},
     \vert \bar{B}^{0}\rangle=\begin{pmatrix}
    0\\1\\0
    \end{pmatrix},
     \vert 0\rangle=\begin{pmatrix}
    0\\0\\1
    \end{pmatrix}.
\end{equation}
Here $B^0$ stands for $B^0_d~ (B^0_s)$ mesons. Further, $\vert B^{0}\rangle$, $\vert \bar{B}^{0}\rangle$ stand for the flavor eigenstates. Similar basis vectors span the decaying $K^0$ meson as well. The time evolution is given by a family of completely positive trace-preserving maps forming a one-parameter dynamical semi-group and is characterized by the operator-sum representation 
\begin{equation}
\rho(t)=\sum_i K_i(t)\rho_0 K_{i}^{\dag}(t).
\label{eq-Kraus_ops}
\end{equation}
The Kraus operators in Eq. (\ref{eq-Kraus_ops}) can be seen in, for example, Ref. [18]. The time evolution is given by a one-parameter dynamical semi-group.
Using this, for example, on a meson initially in the state $\rho_{B^0}(0)=\vert B^{0}\rangle\langle B^{0}\vert$, the evolved state $\rho_{B^0}(t)$ is given by
\begin{widetext}
\begin{equation}\label{statesinglemeson}
\small
    \rho_{B^0}(t)=\frac{1}{2}e^{-\Gamma t}
    \begin{pmatrix}
    \cosh(\frac{\Delta\Gamma t}{2}) +e^{-\lambda t}\cos(\Delta m t) &(\frac{q}{p})^* (-\sin(\Delta m t)-i e^{-\lambda t}\sin(\Delta m t))&0\\
    (\frac{q}{p}) (-\sin(\Delta m t)+i e^{-\lambda t}\sin(\Delta m t)&  \vert\frac{q}{p}\vert^2\cosh(\frac{\Delta\Gamma t}{2})-e^{-\lambda t}\cos(\Delta m t) &0\\
    0&0&2\left(e^{\Gamma t}-\left(1 + \left|\frac{q}{p}\right|^2\right)\cosh(\frac{\Delta\Gamma t}{2})\right)
    \end{pmatrix}.
\end{equation}
\begin{figure}[h]
    \centering
\includegraphics[height=65mm,width=0.45\textwidth]{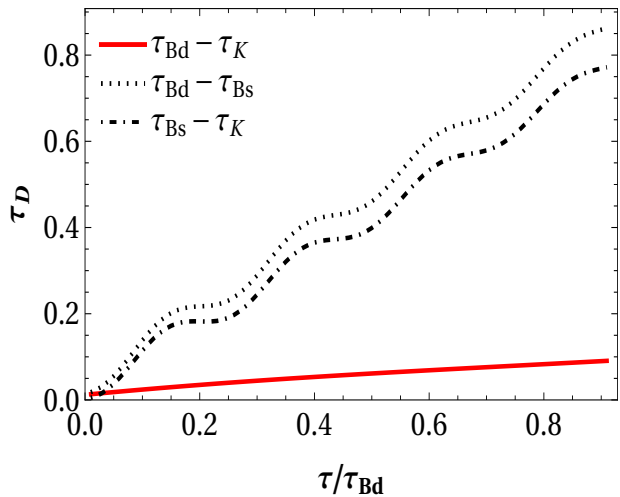}
\caption{Difference in QMTSL for the decoherence-free evolution of  $K^0$ and $B^0_{d/s}$ mesons.}\label{single_m_d}
\end{figure}

\end{widetext}
Here $\Delta \Gamma=\frac{\Gamma_L-\Gamma_H}{2}$ is the difference in the decay width of light, and heavy mesons, $\Gamma=\frac{\Gamma_L+\Gamma_H}{2}$ is the average decay width,  $\Delta m= \frac{m_L-m_H}{2}$ and $m=\frac{m_L+m_H}{2}$ with $m_L$ and $m_H$ being the masses of the light and heavy mesons, respectively. Also, $\lambda$ corresponds to the decoherence parameter.  With the suitable change in notation, the same holds for the $K^0$ meson as well. Here $p$ and $q$ are the coefficients connecting the mass eigenstates $\{|B_L\rangle, |B_H \rangle \}$ to the flavour eigenstates $\{\vert B^{0}\rangle, \vert \bar{B}^{0}\rangle \}$ and are due to CP violation in the flavour oscillations. In particular, $|B_L\rangle = p|B^0\rangle + q|\bar{B}^0\rangle$, $|B_H\rangle = p|B^0\rangle - q\bar{B}^0\rangle$, with $|p|^2 + |q|^2 = 1$. 

 \begin{widetext}
 The linear entropy and coherence, without CP-violation (for ease in analysis), can be obtained by using Eq. (\ref{statesinglemeson}) (without the $CP$ violating terms) in Eqs. (\ref{linearentropy}) and (\ref{coh}), respectively and are 
\begin{eqnarray}
S_{L}(\rho_{B^0}(t))=\frac{3}{4} e^{-2  (\Gamma +\lambda )t} \left(4 e^{ (\Gamma +2 \lambda )t}
   \cosh \left(\frac{\Delta \Gamma  t}{2}\right)-2 e^{2 \lambda  t} \cosh
   (\Delta \Gamma  t)-e^{2 \lambda  t}-1\right)
\end{eqnarray}
\begin{equation}
    Cl_{1} (\rho_{B^0}(t))=\frac{1}{2}e^{-  \Gamma t}(\vert i e^{\lambda t}\sin( \Delta m t)-\sinh(\frac{ \Delta\Gamma t}{2})\vert+\vert i e^{\lambda t}\sin( \Delta m t)+\sinh(\frac{\Delta\Gamma t}{2})\vert).
\end{equation}
 \end{widetext}

\subsection{Correlated mesons}

The initial flavor-space correlated  neutral meson state of the $M\Bar{M}$ meson system $(M=K^0,B^0_d,B^0_s)$ is
\begin{equation}
    \vert\psi\rangle=\frac{1}{\sqrt{2}}(\vert M \Bar{M}\rangle-\vert\Bar{M}M\rangle).\label{correlatedinitl}
\end{equation}
For the $B^0$ meson, this happens due to the decay of the $\Upsilon$ resonance followed by hadronization into a $B^0 \bar{B}^0$ pair. The same holds for the $K^0$ meson, with the $\Upsilon$ being replaced by the $\phi$ meson. By projecting onto the appropriate Hilbert space \cite{banerjee2016quantum,sin2beta}, the dynamics can be shown to evolve the state (\ref{correlatedinitl}) to
\begin{equation} \label{correlatedstate}
\rho(t)=\chi
    \begin{pmatrix}
    \vert s\vert^4 a_- &0 &0 & -s^2 a_-\\
    0& a_+ & -a_+ & 0\\
    0& -a_+&a_+& 0\\
    -s^{*2} a_-&0&0&a_-
    \end{pmatrix},
\end{equation}
where $a_{\pm}=(e^{2\lambda t}\pm 1)(1 \pm \delta_{L})$, $\chi=\frac{(1-\delta_{L})}{4 (e^{2\lambda t}-\delta_{L}^2)}$, 
$\delta_{L}=\frac{2\textrm{Re}(\epsilon)}{1+\vert\epsilon\vert^2}$ and $s=\frac{1+\epsilon}{1-\epsilon}$. 
Here $\epsilon$ is a small CP-violating parameter and is of order $∼ 10^{-3}$ for $K^0$ mesons and $∼ 10^{-5}$
for $B^0_{d,s}$ mesons while $\lambda$ is the decoherence parameter.

Using Eq. (\ref{correlatedstate}) in Eq. (\ref{linearentropy}), the linear entropy of the time-dependent correlated neutral meson state can be shown to be
\begin{equation}
    S_L(\rho)=\frac{4}{3}P_{s}(1-\chi(4 a_+^2+a_-^2(1+s^4)^2)).
\end{equation}
Also, the  $l_1$ norm of coherence is obtained by using Eq. (\ref{correlatedstate}) in Eq. (\ref{coh}) as
\begin{equation}
    C_{l1}(\rho)=2 P_{s}(\vert\chi a_+\vert+\vert\chi a_- s^2\vert),
\end{equation}
where the survival probability $P_s=e^{-2 \Gamma t}\frac{(1-\delta_L^2 e^{-2\lambda t})}{1-\delta _{L}^2}$, and $\Gamma$ is the meson decay width.


\begin{widetext}

\begin{figure} [!htb]
\includegraphics[height=65mm,width=0.45\textwidth]{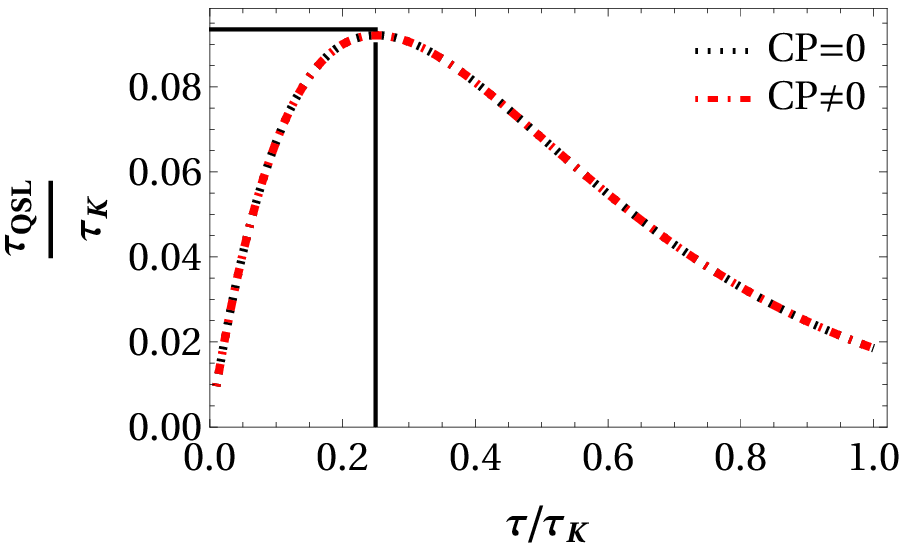}
 \includegraphics[height=65mm,width=0.45\textwidth]{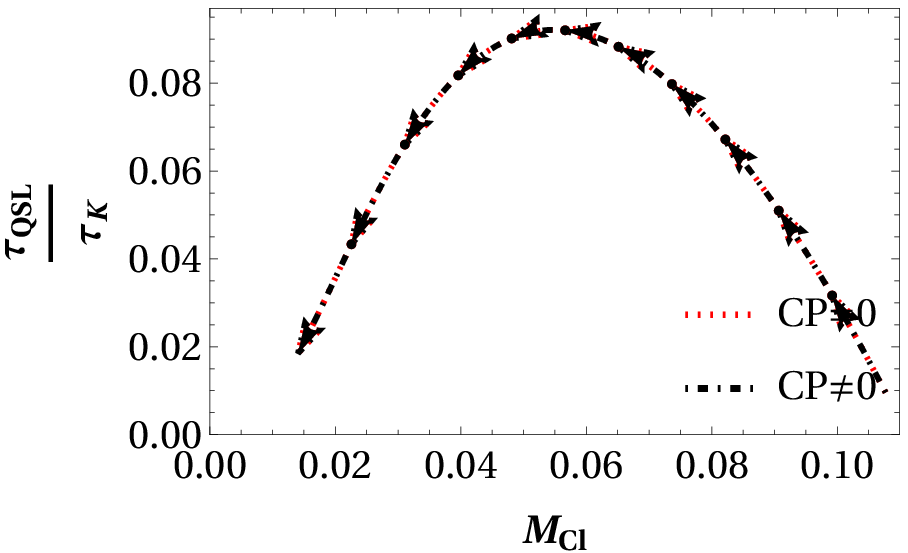}
    \caption{$\tau_{QSL}$ in terms of Bures measure for the decay of the correlated $K^0$ meson system (left subplot), the mean
lifetime is $\tau_{K^0}=1.7889 \times 10^{-10} s$, $\Gamma=5.59\times 10^9 s^{-1}$, $\Delta\Gamma=1.1174\times 10^{10} s^{-1}$, $\Delta m=5.302\times 10^9 s^{-1}$, $m=7.532\times 10^{23} s^{-1}$, $\lambda=2.0\times10^8 s^{-1}$. The impact of the coherence-mixedness trade-off on QMTSL is given in the right subplot.}
    \label{k_correlated}
\end{figure}
\begin{figure}[!htb]
    \centering
\includegraphics[height=65mm,width=0.45\textwidth]{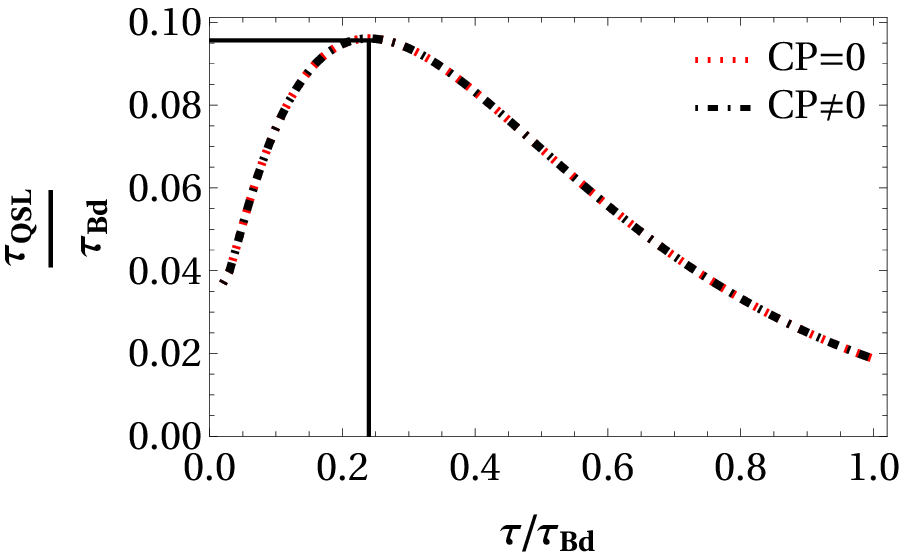}
\includegraphics[height=65mm,width=0.45\textwidth]{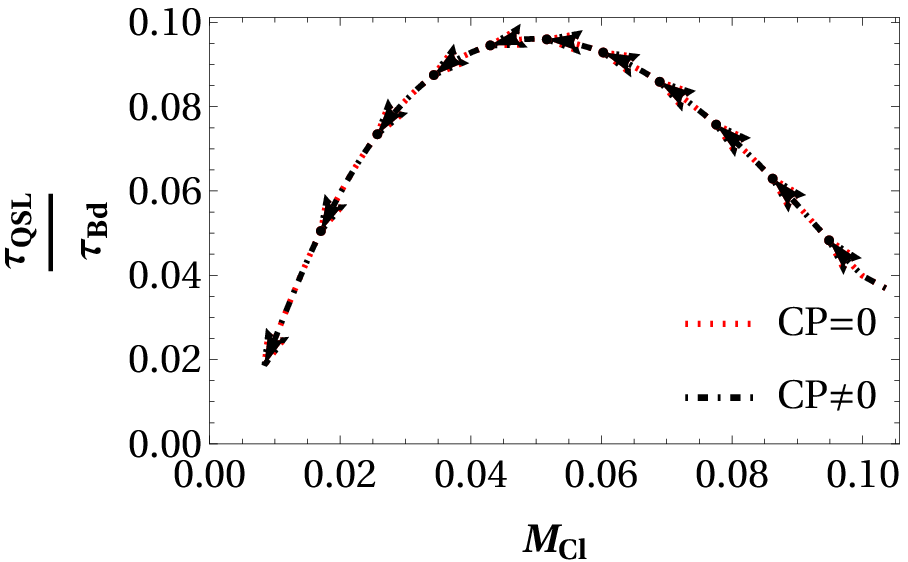}
\caption{QMTSL in terms of Bures measure for the decay of the correlated $B^0_d$ meson (left subplot), the mean
lifetime is $\tau_{B^0_d}=1.518 \times 10^{-12} s$, $\Gamma=6.58\times 10^{11} s^{-1}$, $\Delta\Gamma=0~ s^{-1}$, $\Delta m=0.5064\times 10^{12} s^{-1}$, $m=7.9917\times 10^{24} s^{-1}$, $\lambda=0.012\times10^{12} s^{-1}$. The impact of the coherence-mixedness trade-off on $\tau_{QSl}$  is given in the right subplot.}
    \label{bd_correlated}
\end{figure}

\begin{figure}[!htb]
    \centering
    \includegraphics[height=65mm,width=0.45\textwidth]{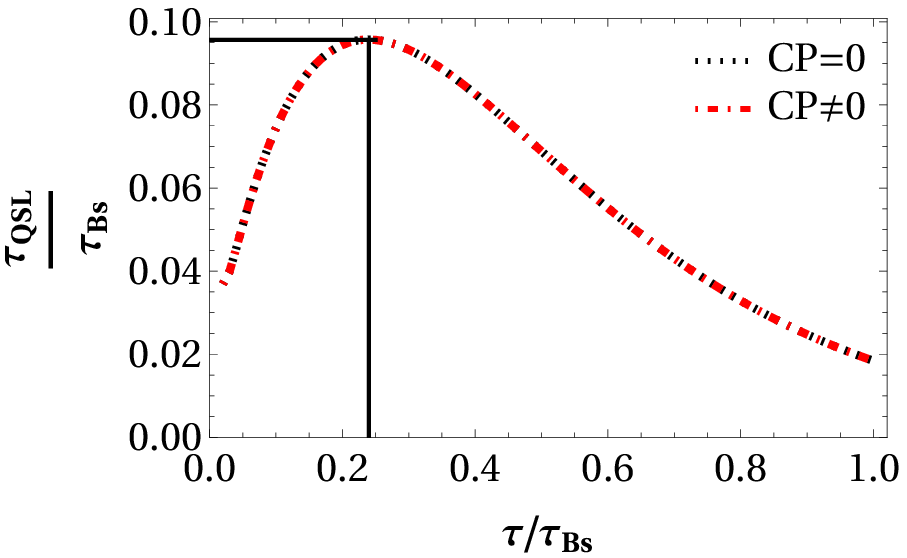}
\includegraphics[height=65mm,width=0.45\textwidth]{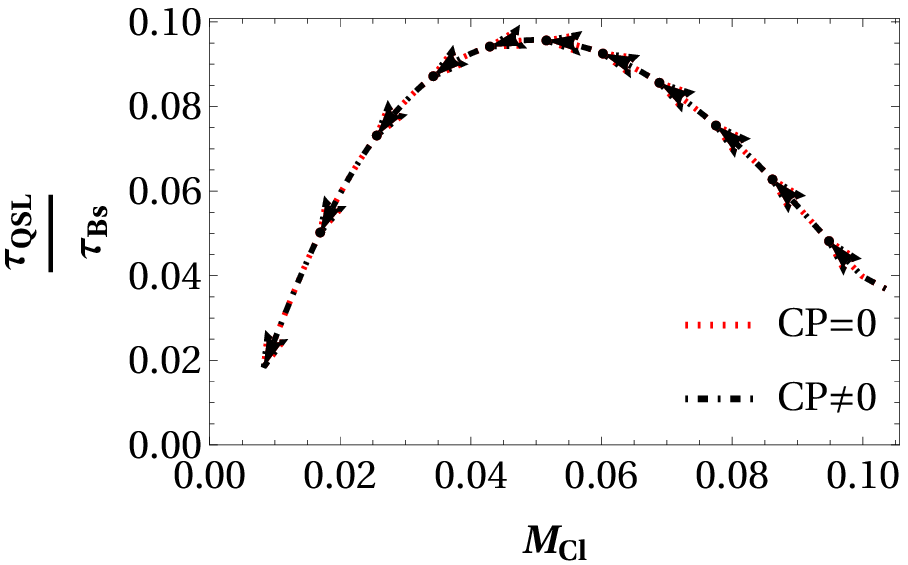}
    \caption{$\tau_{QSL}$ in terms of Bures measure for the decay of the correlated  $B^0_s$ meson (left subplot), the mean
lifetime is $\tau_{B^0_s}=1.509 \times 10^{-12} s$, $\Gamma=0.6645\times 10^{12} s^{-1}$, $\Delta\Gamma=0.086\times 10^{12} s^{-1}$, $\Delta m=17.757\times 10^{12} s^{-1}$, $m=8.123\times 10^{24} s^{-1}$, $\lambda=0.012\times10^{12} s^{-1}$. The impact of the coherence-mixedness trade-off on $\tau_{QSl}$ is given in the right subplot.}
    \label{bs_correlated}
\end{figure}
\end{widetext}

\begin{figure}
    \centering
 \includegraphics[height=65mm,width=0.45\textwidth]{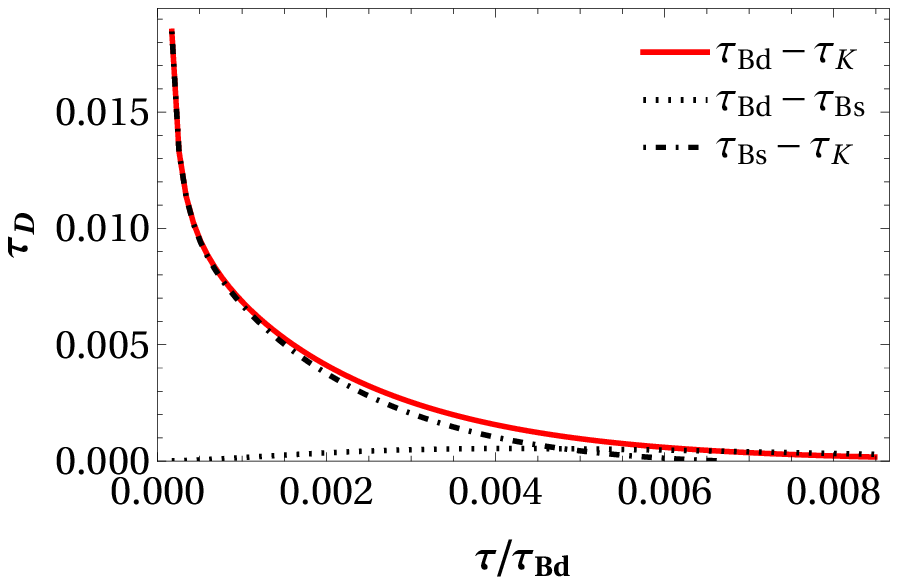}  
 \caption{Difference in the minimum time required for the  evolution of correlated  $K^0$, $B^0_{d}$ and $B^0_{s}$ mesons, with decoherence.}
    \label{ent_difference}
\end{figure}



\section{Results and Discussions: quantum-mechanical time-evolution speed limit}\label{results}

Figures ~\ref{k_meson}, \ref{bd_meson} and \ref{bs_meson} depict $\tau_{QSL}$ (Eq. (\ref{spdlmt})) for neutral $K^0$, $B^0_d$ and $B^0_s$ mesons, along with their antiparticles, plotted with respect to the evolution time  $\tau$. Both $\tau_{QSL}$ and $\tau$ are made dimensionless by scaling with respect to the lifetime of the corresponding mesons. This is pertinent as the mesons are decaying systems and would have ceased to exist after their corresponding lifetime. Generally, the actual time required for a quantum system to evolve from one state to another will be longer than the  QMTSL, which is evident from the figures. It is clear from Fig.~\ref{bs_meson} the  $\tau_{QSL}$ for the decay of   $B^0_s$ is much less than its lifetime compared to other mesons.  The figures also bring out the effect of coherence and mixing, brought by the quantity  $M_{cl}$ (Eq. (\ref{cohmixing})), on the $\tau_{QSL}$. As observed, the $\tau_{QSL}$ and $M_{cl}$ are not able to distinguish between  the mesons and their anti-particles, denoted by $\bar{K}^0$, $\bar{B}^0_d$ and $\bar{B}^0_s$. The insets of the plots highlight the difference caused by taking decoherence, parameterized by $\lambda$, into account. This can be seen to be of the order $10^{-3}$ for $K^0$ and $B^0_d$ mesons, and of the order of $10^{-4}$ for $B^0_{s}$ mesons. The QMTSL increases with the evolution time $\tau$ for $K^0$ and $B^0_d$ mesons. This is a signature of the underlying semi-group nature of the evolution \cite{caban2005unstable}. It is consistent with the corresponding behaviour under other Markovian evolutions \cite{paulson2021hierarchy}, generated by a quantum dynamical semi-group \cite{breuerpettrucione, sbbook}. Different from the $K^0$ and $B^0_d$ mesons, the $\tau_{QSL}$ for $B^0_s$ mesons exhibit a wiggling pattern, which could be attributed to the parameter values it takes. This is in accord with a similar behaviour seen for the Leggett-Garg quantity for $B^0_s$ mesons \cite{LGMesons}. The mass term $\Delta m$ for $B^0_s$ meson, which plays the role of frequency, is more than 35 times the corresponding value for the 
$B^0_d$ meson. This would account for the difference in the $\tau_{QSL}$ between the $B^0_d$ and the $B^0_s$ mesons. Figure (\ref{single_m_d}) brings out the difference in QMTSL for the evolution of  $K^0$ and $B^0_{d/s}$ mesons, with respect to evolution time $\tau$, scaled with the lifetime of the $B^0_d$ meson, in the absence of decoherence. The wiggle in the curves showing the difference with the $B^0_s$ meson can be attributed to the above-mentioned feature of these mesons.

From Figs.~\ref{k_correlated}, \ref{bd_correlated} and \ref{bs_correlated}, for correlated mesons under decoherence, it can be seen that as the evolution time increases to around $0.24~ \tau_{lifetime}$,  $\tau_{QSL}$ increases, and then falls. This implies that for an evolution time of up to approximately $1/4^ {\textrm{th}}$ of the lifetime, the evolution of correlated mesons slows down, and beyond this, as it decays, the evolution is sped up. This is consistent with the behaviour of coherence-mixing balance  depicted in the right panel of the figures. Here, it can be seen that, following the direction indicated by the arrows, as the $M_{cl}$ quantity decreases, $\tau_{QSL}$ first increases and then falls.  The difference in speed limit ($\tau_D$) of evolution for different correlated mesons with respect to evolution time $\tau$, taking decoherence into account, and scaled with the lifetime of the $B^0_d$ meson, are depicted in Fig.~\ref{ent_difference}. These curves highlight that $\tau_D$ for $B^0_d$, $K^0$ and $B^0_s$, $K^0$ are of the order $10^{-3}$, whereas for $B^0_d$, $B^0_s$, it is negligible. This difference approaches zero very quickly, within the $B^0_d$ meson lifetime. We also observe that the maximum speed limit time for correlated mesons is lesser than that for single mesons. This shows that correlated mesons evolve faster than single uncorrelated mesons. This is consistent with observations made in different systems that quantum correlations can speed up the evolution, see for example \cite{sbcentralspinspeed}. Further, the impact of $CP$ violation on the dynamics of $\tau_{QSL}$ is negligible for the mesons studied.

\section{Conclusion}\label{conclusion}
In this work, we investigated the quantum-mechanical time-evolution speed limit (QMTSL) in  neutral $K$ and $B$ mesons, both single as well as correlated, within the framework of open quantum systems. The aim was to better understand the dynamics of these unstable subatomic systems by investigating the impact of decoherence and CP violation on the QMTSL. 
The impact of decoherence is seen to be minimal. The coherence--mixing, a crucial feature of the open system evolution of the underlying quantum systems (here,  the neutral mesons), has an important influence on the  QMTSL. Both the QMTSL and coherence--mixing balance are not able to distinguish single neutral mesons and their~anti-particles.

The results showed that for single mesons, the~QMTSL increased with the evolution time, which is consistent with the semi-group nature of the evolution. However, for~correlated mesons, the evolution slowed down for a quarter of the meson's lifetime and then sped up, reflecting the balance between decoherence and quantum correlations in the evolution. The~difference in speed limit $\tau_{D}$ of evolution for correlated $B_d^0$, $K^0$ and $B_s^0$, $K^0$ were seen to be of the same order, whereas they were negligible for $B_s^0$, $K^0$. The results also showed that correlated mesons evolve faster than single, uncorrelated mesons, suggesting that quantum correlations can speed up evolution.

\section*{Data availability statement}
The manuscript has no associated data.
\section*{Acknowledgement}
SB acknowledges the support from the Interdisciplinary Cyber-Physical Systems (ICPS) programme of the Department of Science and Technology (DST), India, Grant No.: DST/ICPS/QuST/Theme-1/2019/6. SB also acknowledges support from the Interdisciplinary Research Platform (IDRP) on Quantum Information and Computation (QIC) at IIT Jodhpur.

 
\end{document}